\documentclass[11pt,a4paper]{article}
\pdfoutput=1
\usepackage{jcappub}

\usepackage{color}

\title{Particle ejection during mergers of dark matter halos}

\author[1]{Isabella P. Carucci,}
\author[1]{Martin Sparre,}
\author[1]{Steen H. Hansen,}
\affiliation[1]{Dark Cosmology Centre, Niels Bohr Institute,\\
University of Copenhagen, Juliane Maries Vej 30, 2100 Copenhagen, Denmark}
\author[2]{and Michael Joyce}
\affiliation[2]{Laboratoire de Physique Nucl\'eaire et Hautes \'Energies, \\
Universit\'e Pierre et Marie Curie - Paris 6, CNRS IN2P3 UMR 7585, \\  
4 Place Jussieu, 75752 Paris Cedex 05, France}

\abstract{ Dark matter halos are built from accretion and merging. During merging some of the dark matter particles may be ejecteted with velocities higher than the escape velo\-ci\-ty. We use both N-body simulations and single-particle smooth-field simulations to demonstrate that rapid changes to the mean field potential are responsible for such ejection, and in particular that dynamical friction plays no significant role in it. Studying a range of minor mergers, we find that typically between $5-15\%$ of the particles from the smaller of the two merging structures are ejected. We also find that the ejected particles originate essentially from the small halo, and more specifically are particles in the small halo which pass later through the region in which the merging occurs.
}

\keywords{dark matter: halos, techniques: numerical}

\begin{document}
\maketitle

\section{Introduction}

Observations suggest that structure formation in the Universe proceeds
hierarchically, with the smallest structures collapsing first and then
later merging to form larger structures. An important property of mergers 
is that if two halos with identical density profiles merge, the merger remnant
will have a density profile with the same inner and outer slopes as the initial 
profiles \citep{2006ApJ...641..647K}. It has been proposed that this behaviour may be understood from considerations about mixing of collisionless
systems \citep{Dehnen2005}. Further merger remnants have been found to be
triaxial with a major axis along the collision axis~\citep{2004MNRAS.354..522M}, 
and the velocity anisotropy profiles
are typically radial, in the sense $\sigma_{\rm r}^2 \ge \sigma_{\rm tan}^2$,
whether calculated in spherical or
elliptical bins. The velocity anisotropy profiles of merger remnants
are, however, not simple functions; they are asymmetric in the sense
that a very different behaviour is seen along different axes
\citep{2012JCAP...07..042S}. It has been pointed out~\citep{2013arXiv1303.2056W}
that the velocity ellipsoids of a halo are typically
aligned with the major axis, and it means that the velocity anisotropy
parameter gives a misleading description of triaxial halos.

The relaxation and mixing processes in collisionless mergers have been
examined in \citep{2007ApJ...658..731V}, where it was found that mixing of the
6-dimensional phase space distribution function mainly occurs during
the tidal shocking arising when the center of the merging halos pass
through each other, and that about
40 \% of the particles
from the merging halos are located outside the virial radius of the
remnant. In controlled numerical galaxy collisions it has been
known for a long time that some particles are ejected with positive
energies (see e.g. \citep{hernquist92}).  In cosmological simulations it has
been found that unbound particles are abundant in halos which have
recently undergone a major merger~\citep{2012arXiv1208.0334B}.

Why particles are ejected in the course of a merger is straightforward 
to understand on energetic grounds. When a small structure is engulfed by 
a larger one, it will be ripped apart dynamically, and a new equilibrium 
state will be reached. At the end of this equilibration, the virial 
theorem must hold for the bound particles, i.e., 
\begin{equation}
2 K + W = 0 \, ,
\end{equation}
where $K$  and $W$ are the total kinetic and potential energy of the 
bound particles (with $K$ calculated relative to their centre of mass). 
The total energy of the virialized particles is thus $E_b=-K$.
For the case of two virialized structures initially 
very far from one another, the total energy can be written as 
$E= -K_1-K_2+K_{CM}$ where $K_1$ and  $K_2$ are the 
initial kinetic energies of the two systems (relative to their centre of 
masses) and $K_{CM}$ is the kinetic energy of the centre of
masses of the two halos when they are far apart.  
By energy conservation we therefore have that particles must be
ejected if $K_{CM}$ in the centre of mass frame, i.e., 
the kinetic energy of the initial relative motion, is sufficiently 
large to make the total initial energy positive: dark matter,
unlike baryons which can radiate, have no other way of
discarding the excess energy in the system.  In the case of
cold dark matter, such relative motion is typically small, and 
we will in fact assume it to be zero, i.e., the structures start
far apart without initial relative motion.
In this case, from a global energetic point of
view, ejection {\it may or may not} take place. However, in
the case of a small structure which encounters a very
large structure it is not difficult to see why such ejection
{\it does} in fact typically take place, and why the ejected
particles are (as we will verify in detail below) those 
in the small structure. Neglecting completely the 
potential energy due to the small structure, we
can consider the particles belonging to it as 
unbound particles falling into the fixed potential 
of the large structure. The condition of 
virialization in this fixed potential then imposes that 
energy must be ejected: 
for example, if two 
particles arrive from far away 
with zero energy, and virialize
in this way,  the combination of the virial
condition and energy conservation
would then give
$2(K+k_a+k_b) + (W+w_a+w_b)=
k_a+k_b=0$, where $k_a$ ($w_a$)
and $k_b$ ($w_b$) are the average
kinetic (potential) energies of the 
particles,  a condition which cannot
be satisfied. The system, however, can reach
a virial equilibrium without any 
signification modification of the large 
structure by ejecting particles carrying
away excess energy. 
From a dynamical point of view, the particles in 
the  small structure are much less bound and the potential 
fluctuations induced by the merger are of order 
the initial potential energy of this structure, 
precisely of the order required to eject them.

In this paper we perform numerical simulations to study 
and characterize the ejection of particles in mergers and 
to understand the mechanism responsible for this ejection. 
We use $N$-body simulations to study the ejection
of particles in minor mergers, and we use single particle simulations
in smooth potentials to demonstrate that the ejection mechanism is
indeed a mean-field effect. Particle (and energy) ejection during the 
process of violent relaxation has been previously discussed
at length in~\cite{2009MNRAS.397..775J}, and studied in detail with $N$-body
simulations for the case of cold quasi-spherical initial conditions.
Typically about $15\%$ of the particles are found to ejected in 
this case, and it is shown that these are particles initially in
the outer shells which ``arrive late" at turnaround and pick up a 
positive energy kick as they travel 
in the potential sourced
by the re-expanding 
potential of the bulk of the particles. Despite the quite different 
initial conditions studied here, we will see that the 
mechanisms at play in the ejection are in fact very similar.

\section{$N$-body simulations}
\label{sub:nbody}

In this section we will describe our $N$-body simulations of minor
mergers, which have been performed using 
the \texttt{GADGET-2} code \citep{2005MNRAS.364.1105S,springel_gadget}.
All the simulations were pure dark matter simulations, where dark matter 
is modelled as collisionless particles, and in a non-expanding universe
with open boundary conditions.

\begin{figure}
\includegraphics[scale=0.8]{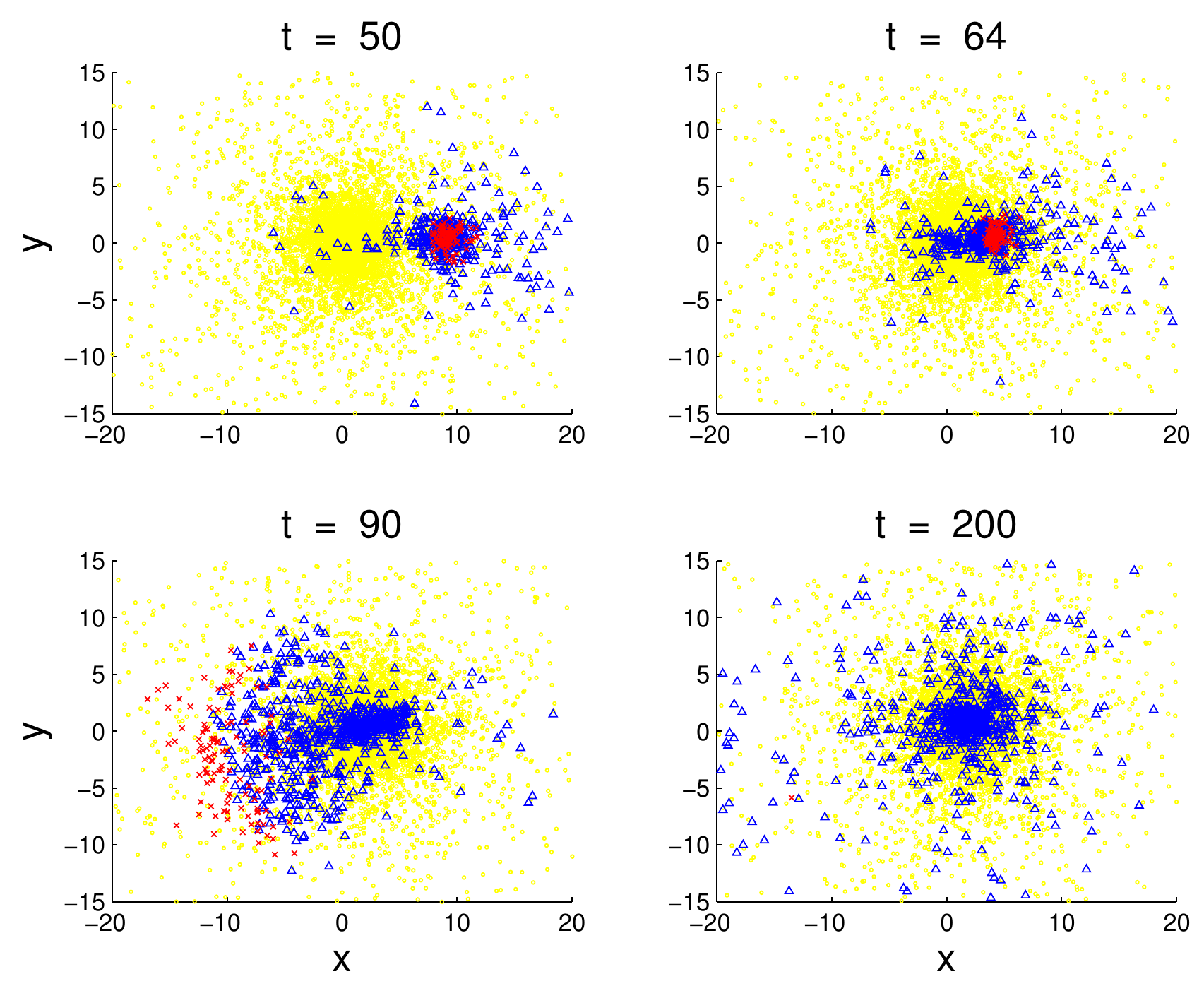}
\caption{The positions in $x$ and $y$ of the particles at different
  time steps of the simulation. The yellow dots are particles
  belonging to the major halo, the blue triangles are the minor halo
  particles that stays bound throughout the simulation, and the red
  stars are the particles which get ejected. At $t=64$, it can be
    seen that the ejected particles from the minor halo arrive at the
    plane of the center of the major halo slightly later than the center of the
    minor halo does.}
\label{fig:snaps}
\end{figure}

\subsection{Simulation setup}
\label{sub:setup1}

We study mergers between halos with a mass ratio of 1:10,
representative of typical mergers encountered in cosmological
simulations. In our simulations we have $10^5$ particles in the 
big halo and $10^4$ particles in the small halo.
We work in units in which the gravitational
constant $G$ is unity. Further we take both the mass of the larger halo,
$M_1$, and its scale radius, $r_\text{s}$, to be unity.  All particles have
the same mass. 
 We use a gravitational softening given by $0.023$ in the
\texttt{GADGET-2} parameter file,  and force and time integration 
accuracies are fixed by  \texttt{ErrTolForceAcc} $=0.005$ and 
\texttt{ErrTolIntAccuracy}  $=0.025$, respectively. We will discuss
further below tests we have performed on the stability of our results
to variation of these parameters.

The positions and the velocities of the particles are chosen so that
each structure, treated as an isolated system, is in a steady state
in virial equilibrium. Positions are assigned from the
Hernquist mass profile \citep{1990ApJ...356..359H},
\begin{align*}
\rho (r) = \frac{1}{r/r_\text{s}} \frac{\rho_0}{(1+r/r_\text{s})^3},
\end{align*}
and the velocities are selected from a Gaussian PDF with the initial
isotropic velocity dispersion derived from the Jeans equation.  The
initial velocities are truncated at $0.95 v_\text{esc}$.  We ran
simulations with different scale lengths, $r_{s2}$, of the small halo
(but with fixed mass).

The centres of mass of the two halos are placed at $y=0$ and $z=0$ in
our cartesian coordinate system.  Instead in the $x$ direction we place
the big halo at $0$ and the small at $15$. The latter value is a rough
estimate for the turnaround radius of the big structure
\citep{2006ApJ...645.1001P,2008MNRAS.389..385C}, assuming it
to have a typical concentration of galaxies \citep{2008MNRAS.391.1940M}.  
Having chosen to place the minor halo at the turnaround radius of the major,
we start the simulation start with both halos at rest.
The small structure starts approaching the big one, pulled by
its gravitational attraction.
After the merger we continue the simulation for at least $10$ more
dynamical times, where we define a typical dynamical time from the
circular velocity at $r_4=4\,r_{s1}$, $\tau_\text{dyn}=r_4/v_c(r_4)$.  
When we observe that the total energy of the new structure (without the
ejected particles) is constant, we deduce the system has reached a
new equilibrium.

\subsection{Results}

\subsubsection{Identifying the ejected particles}
\label{sub:example}

Our first simulation uses a scale radius $r_{s2} = 0.3$ for the small
halo. The positions of the particles as a function of time is shown in
figure \ref{fig:snaps}. The red stars in the plot are particles which
start in the minor halo and are ejected during the simulation, where
we label a particle \emph{ejected} if its total energy in the
final snap-shot is positive.

The colour code allows us to follow the ejected particles in red from
the beginning of the run, to distinguish them from the 
other particles belonging initially to the small structure, in blue,
and those belonging to the big structure, in yellow.  We see
that, at $t=50$, the small halo has not yet crossed the big halo core and
the red particles are well mixed with the blue ones. As the
small halo enters the large halo's core at $t=64$, we see that 
the red particles are those which  \textquotedblleft lag
behind\textquotedblright and are the last ones to cross the core. The
next snapshot describes the subsequent ejection, and in the last
one at $t=200$ ($\sim 23$ dynamical times) there are no  longer
red particles within 15 times the scale radius.

\subsubsection{Numerical tests}

As pointed out in the introduction, we expect ejection to happen generically in this kind of merger.
Nevertheless to check that the mass ejection we observe numerically is not the consequence of 
some other effect, notably ill-posed initial conditions (i.e. particles not in equilibrium to begin with),
or accumulated integration errors in the particle orbits, we run additional specific test simulations.

We test first that our results are insensitive to the use of the Gaussian
approximation of the initial velocities. Specifically, we use the
Eddington inversion method \citep{binneytremaine} to set up initial equilibrated
structures (using an implementation which was tested in
\citep{attractor,2012JCAP...10..049}), and we find that merger time and number of
particles ejected are identical within a few percent. Also the
interesting feature of figure~\ref{fig:snaps}, that the ejected particles 
are the ones arriving ``late'' is exactly the same for initial structures created
using the Eddington method.

Figure~\ref{fig:snaps} keeps its characteristics also when we run more accurate simulations,
taking  \texttt{ErrTolForceAcc} $=0.001$ and \texttt{ErrTolIntAccuracy} $=0.005$ (i.e. a factor of
five smaller than in our fiducial simulations). In this case too, merger time and number of 
particles ejected are stable within a few percent.

\subsubsection{Which particles are ejected?}

The mechanism of ejection is in fact simply that of violent
relaxation in general, as originally described by
 \citep{1967MNRAS.136..101L} for stellar systems:
starting from an initial configuration which is far
from dynamical equilibrium, such a system can
relax precisely because,  in the time dependent 
gravitational potential, particles' energies can 
change rapidly (i.e. on mean field time scales).
If the fluctuations of potential are sufficiently 
violent, particles can reach and surpass 
the escape velocity of the system.

\begin{figure}
\includegraphics[scale=0.8]{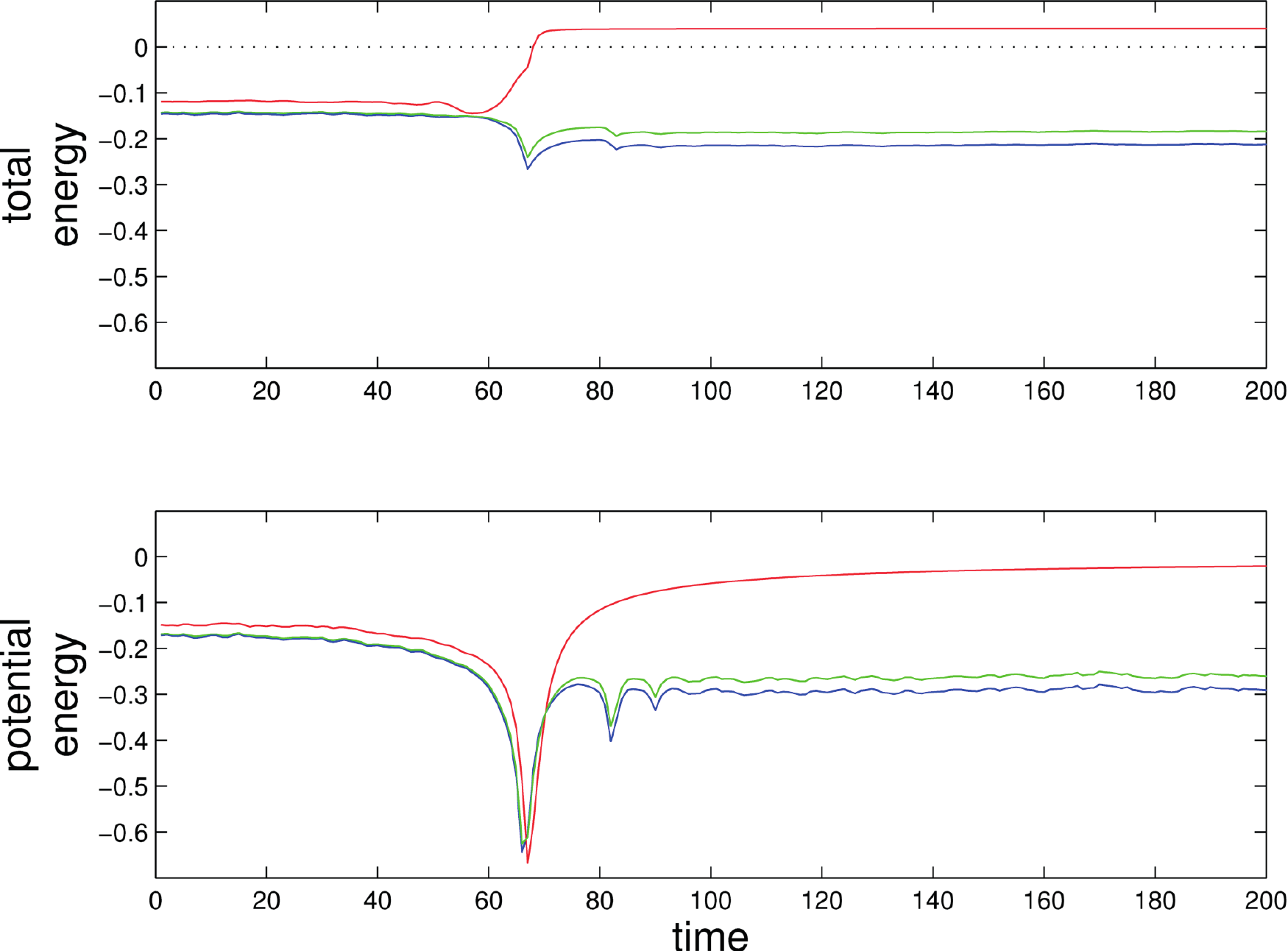}
\caption{Time evolution of the total energy (upper panel) and potential
  energy (lower panel) of all the particles belonging to the small
  halo (green), of the particles which remain bounded (blue), and of 
  those which are ejected (red).}
\label{fig:energies}
\end{figure}

More specifically it is the particles that just happen 
to arrive later in the region at the centre of the merger which
pick up a large positive kick to their energy in a short time as they 
pass through the time-dependent potential well created by the
rest of the mass. To see that this is the case, we plot in
the lower panel of figure \ref{fig:energies} the average potential
energy of the blue (bounded) and red (escaping) particles
as defined in the previous figure, as well as the average
over all particles (in green).  It can be seen clearly that the red particles pass on average through the minimum of the potential well slightly later than the blue ones.

Further from the plot in the upper right panel of figure~\ref{fig:snaps}, it 
can be seen that all the ejected particles were on the right side of the
minor halo at $t=64$. 

\begin{figure}
\includegraphics[scale=0.8]{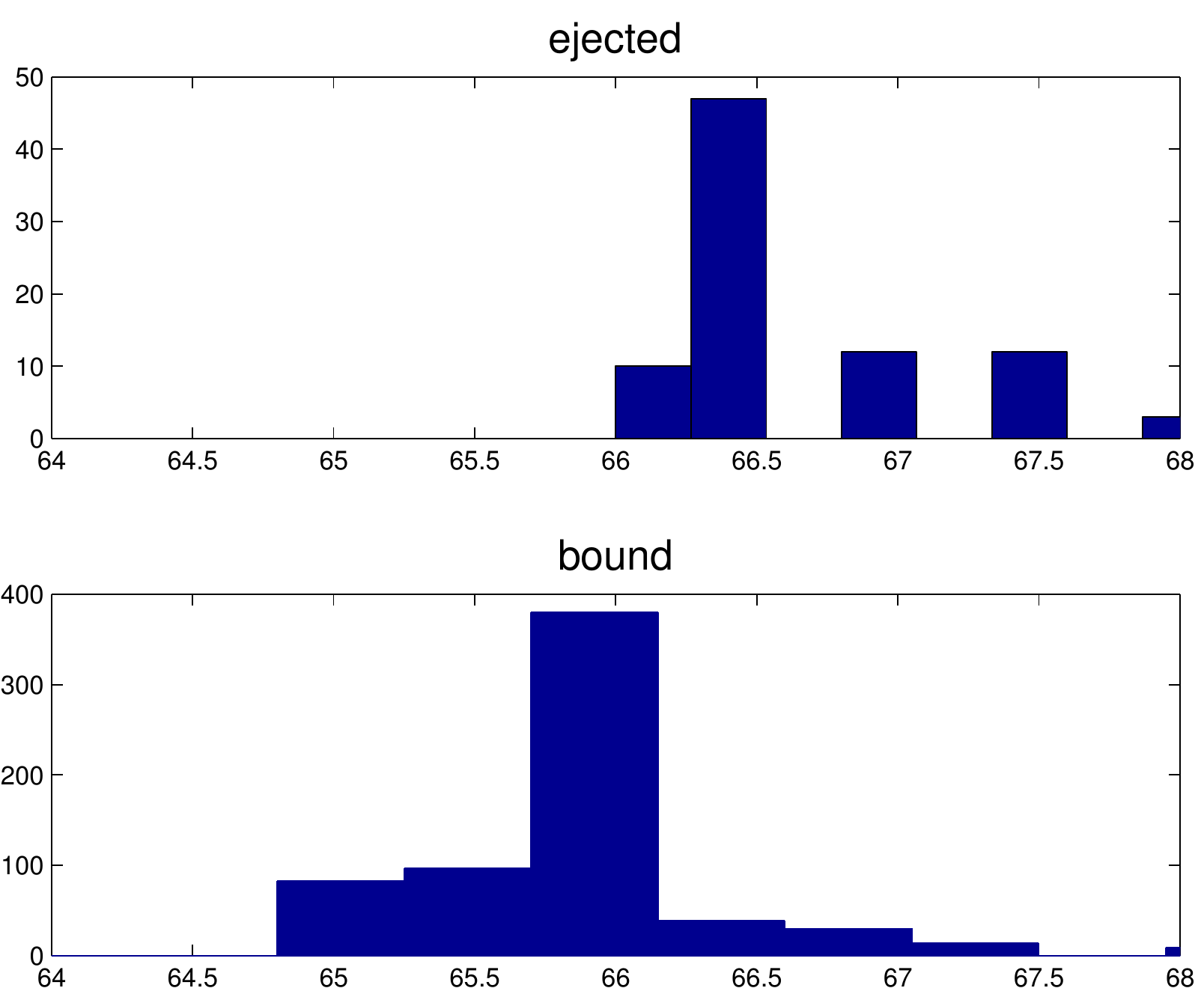}
\caption{The distribution of ``dip-times'' for all the particles which
  are either ejected or bound. The dip-time is defined as the time
  when the potential energy of the particle reaches its minimum.  The
  particles which are ejected are seen statistically to have a later
  dip-time than the particles which are bound.}
\label{fig:diptime}
\end{figure}

In order to show that the ejected particles are the late arrivers, we
follow the orbit of each particle from the small halo, and we find the
time when the potential energy is at a minimum for each particle. For
this figure we include only particles inside 5 times the scale radius.
We can now plot the distribution of these ``dip-in-energy" times for
the particles which will eventually be ejected, and compare this to
the particles which will be bound, and we clearly see the difference
in Figure~\ref{fig:diptime}. We see that the peak of the ejected
particles is later than the peak for the particles which remain bound.

\begin{figure}
\includegraphics[scale=0.8]{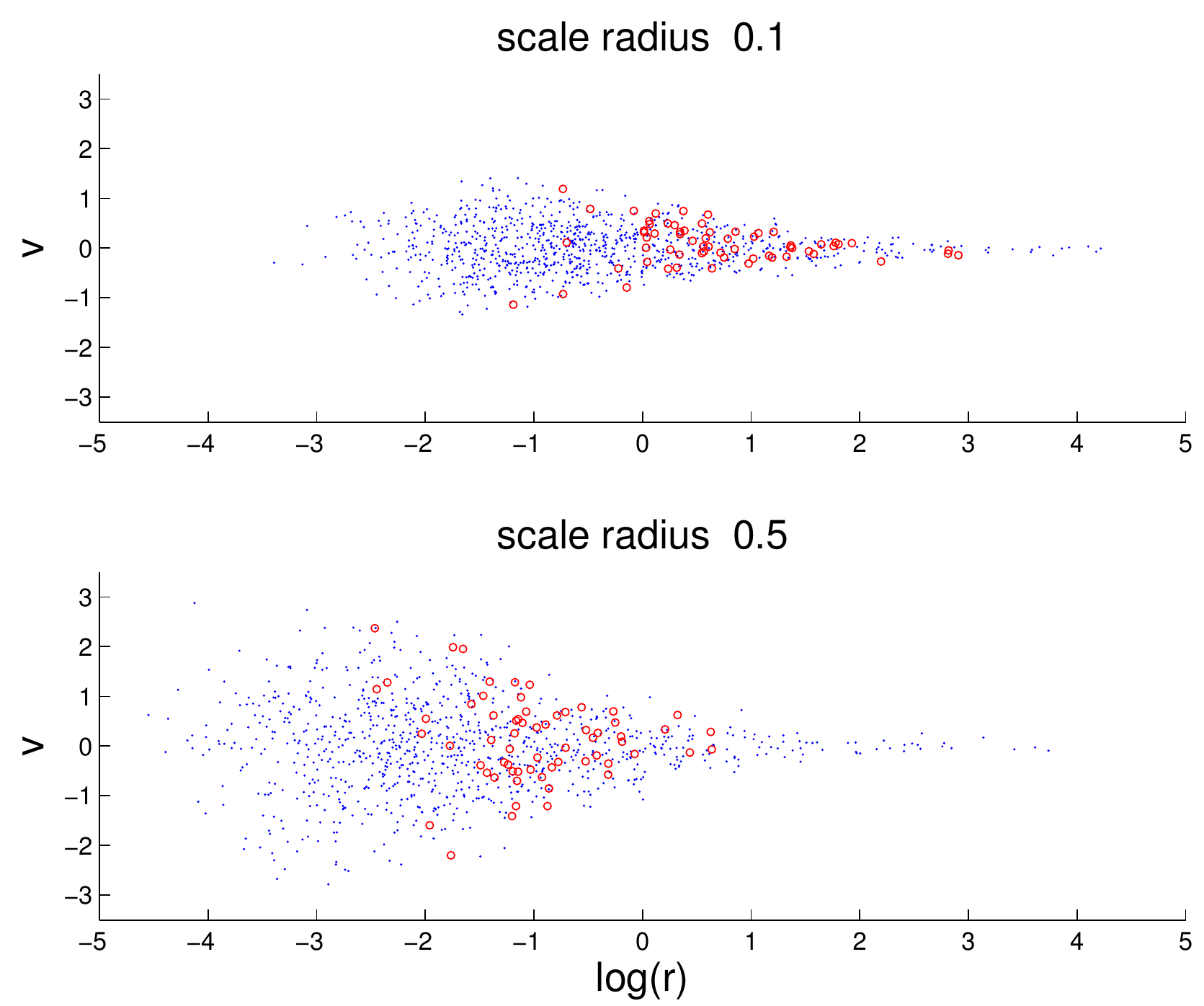}
\caption{Radius $r$ and radial velocity $v$ normalised with the
  related virial quantities for $2$ different scale radii for the
  small structure. The red circles represent the particles which will
  later be ejected, the blue dots are the particles which remain bound.}
\label{fig:vr}
\end{figure}

In figure \ref{fig:vr} we plot the radial velocities and the radii of
the particles in the small halo before the merger happens, for two
different values of $r_\text{s}$. The red circles label the particles that will
subsequently be ejected. They are distributed almost as the other particles: 
in the sense that at any given radius the velocity distribution
of the ejected particles  is similar to that of the particles
which remain bound. Furthermore, the ejected particles 
 are neither the most energetic nor those furthest from the centre 
of the structure. They are not, on the other hand, not the particles 
orbiting at the smallest radii, and hence not the most bound particles.
Further plotting other quantities such as the angular
momentum (modulus and direction) we have not identified 
any particular feature that characterises the particles in question. 
The decisive factor thus  seems to be whether or not a particle 
is falling early or late into the combined potential of the cores 
of the halos.

\subsubsection{Analysis of particle energies}

\begin{figure}
\includegraphics[scale=0.8]{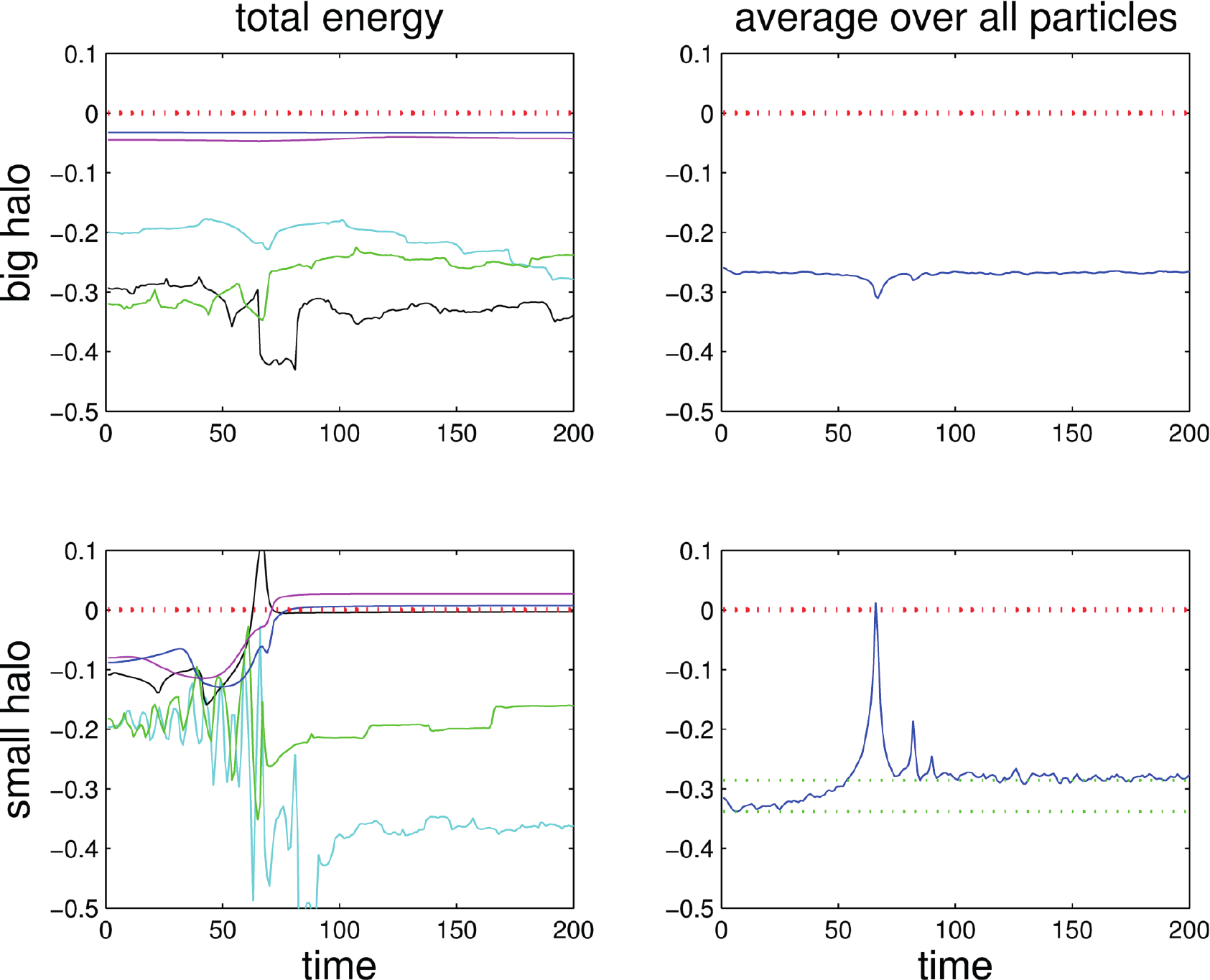}
\caption{Time variation of the total energy.  On the left we see the
  time evolution of five different particles, and on the right the
  average over all the particles. The upper plots are for the big halo, 
  the lower ones for the small halo. The red dotted line shows the zero 
  point of the energy.}
\label{fig:example}
\end{figure}

The particles which are ejected
are those of which the energy is positive after the merger.
Figure \ref{fig:example} shows the evolution of
the energy of a few randomly selected particles. In 
the case of the big structure, its particles' energies 
stay roughly constant in a range of values well
represented by the total energy of the halo 
averaged over the number of its particles.
None of the particles from the big halo are ejected during the merger.
Among the 
chosen particles belonging to the small halo, there are a couple 
for which the total energy becomes positive after the merger, 
roughly in a time window between $t=60$ and $t=70$. 
None of the particles are ejected before $t=50$.
The mean particle energy of the small halo displays a sharp peak 
at the moment of the merger and then falls back to a roughly 
constant value, which is higher than the initial one.
Thus, during a collision in this simulation $11\%$ of the 
particles from the small halo are freed from the
system during the merger, and will never return.

\subsubsection{Mechanism of ejection}
\label{sec:ejection}

The observation that it is the particles coming in later which
are ejected is similar to the case of cold uniform spherical collapse
in  \citep{2009MNRAS.397..775J,sylos}, where the ejected particles were
found to be those starting out in the outer shells. In this case 
close analysis of the particle energies shows that those which
escape pick up the energy leading to their ejection when they 
pass through the potential generated by the bulk of the 
mass which has already turned around and starting 
re-expanding: as the time derivative of the potential
is then positive, they gain energy.

\begin{figure}
\includegraphics[scale=0.8]{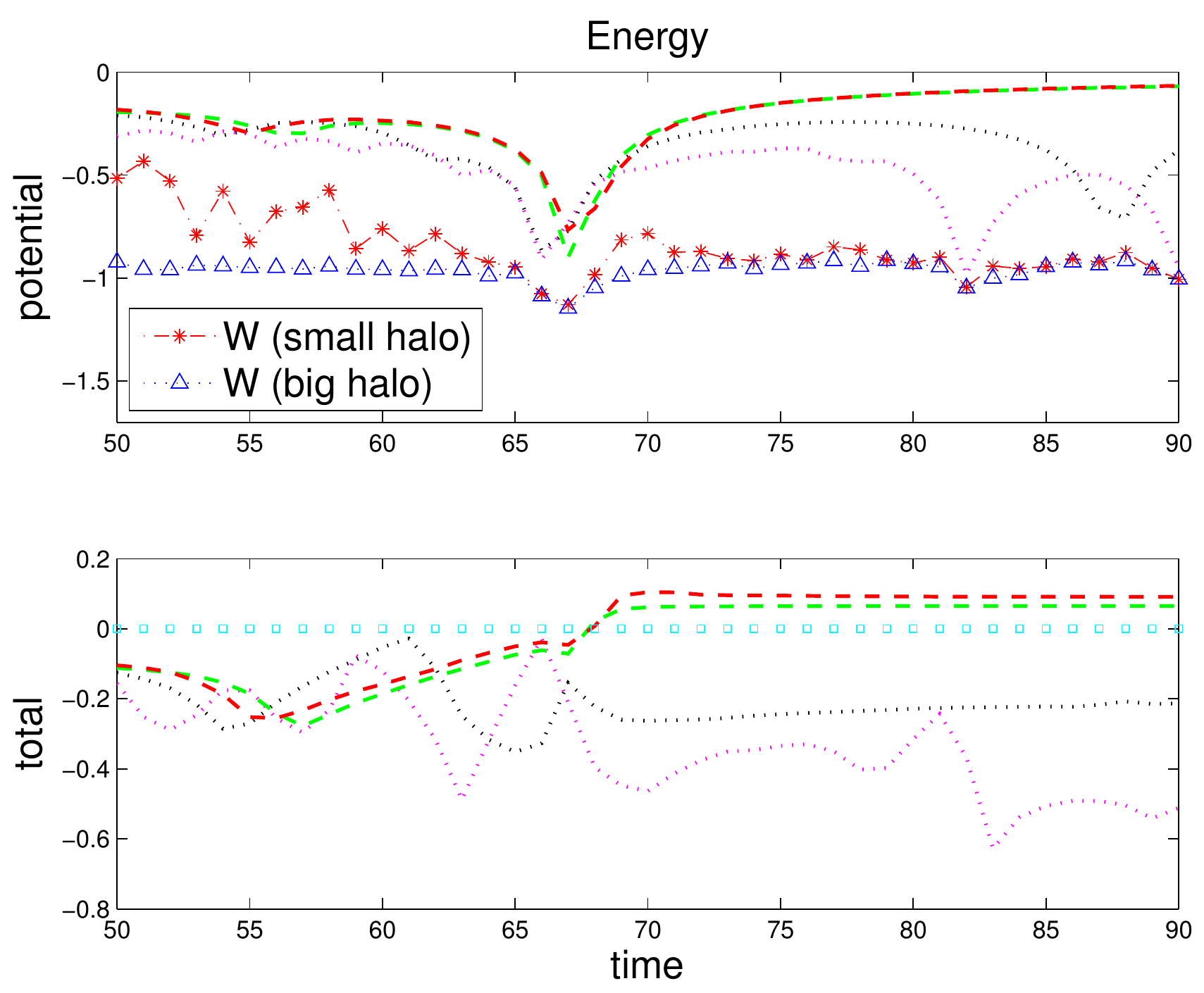}
\caption{Zoom in on the time variation of potential and total energy. The particles
  which will remain bound (blue and purple dotted lines) pass  through
  the central region while the potential due to both the small and large 
  structures (symbols) are deepening (around time step 66). This contrasts with 
  the particles  which will be ejected (red and green dashed lines), which arrive 
  a little later (time step 67) and pass the centre when the potential
  is decaying.}
\label{fig:rdm}
\end{figure}

A detailed view of the time evolution of particle energies here 
reveals that what is happening in the minor merger is very
similar. In figure \ref{fig:rdm} we zoom in on the time steps during
the merging of the small structure. We follow again the evolution of 
five particles, where the green and red (dashed lines) are ejected
particles, whereas the blue and purple (dotted lines) are finally
bounded.  Further we show the energy of the most bound
particle from the small structure (stars), and the same from the big
structure (triangles).  Inspecting the time steps between 62 and
67, we see that the potential of both the small and the large
structures get deeper. This is natural because during the first
passage when the 2 structures overlap for the first time, the
potentials deepen. However, during the time-steps from 67 to 
70 the potential falls back to a smaller absolute value. This is 
just after the first passage of the small structure. After this point, the
potential flattens out close to the final value.

Now, comparing the time of passage of the particles which will 
remain bounded (black and purple dots), we see that they pass the center 
during the deepening of the potential of the structure. The 
particles which are ejected, on the other hand, arrive later, 
during the phase when the potential happens to be 
weakening (i.e. increasing towards less negative values).
As the sign of the time derivative of the mean field
potential is positive the energy of the particles increase,
since the time variation in particle energy
along a trajectory is equal to the time derivative of 
potential energy~\citep{1967MNRAS.136..101L}, $dE/dt = \partial \Phi/ \partial t$.

While we have shown here the time evolution of only two ejected particles, 
we underline that figure 1 shows that of the roughly 1000
ejected particles, essentially all of which arrive late during the
merger as we have described. We note that this effect is analogous 
to the so-called late-time integrated Sachs-Wolfe effect, in which photons
gain energy because of the decay of the potential they are 
traversing.

We note that temporal variation of the mean field potential is 
indeed well known to induce changes in the energy of individual particles, 
potentially even leading to evaporation of particles, e.g. discussed as 
tidal forces \citep{1958ApJ...127...17S}, and as gravitational shocks during merging 
\citep{1972ApJ...176L..51O}. Very large changes in energy are
also known to occur during impulsive tidal shocking 
\citep{2007ApJ...658..731V} which
exactly happens during pericenter passages.

Whereas these previous studies have considered average changes in
energy, we emphasize here that , like in the case of spherical cold collapse
\citep{2009MNRAS.397..775J}, the particles which 
pick up enough energy to be ejected are those specific ones which 
arrive late relative to the time at which the total potential reaches
its minimum. These particles are thus ejected because of a coincidence 
between their individual orbits combined with the potential changes 
during merging.

\subsubsection{What determines the fraction of ejected particles?}

After having considered the phenomenon of ejected particles in mergers,
we proceed by performing simulations with different sizes of the
minor halo. We run different simulations keeping the scale radius of
the big halo set to $r_\text{s} = 1$, while changing the small halo $r_\text{s}$
within the range $0.1 \leq r_{s2} \leq 0.7$. The ratio of the halo
masses is unchanged (equal to $0.1$).

\begin{figure}
\includegraphics[scale=0.8]{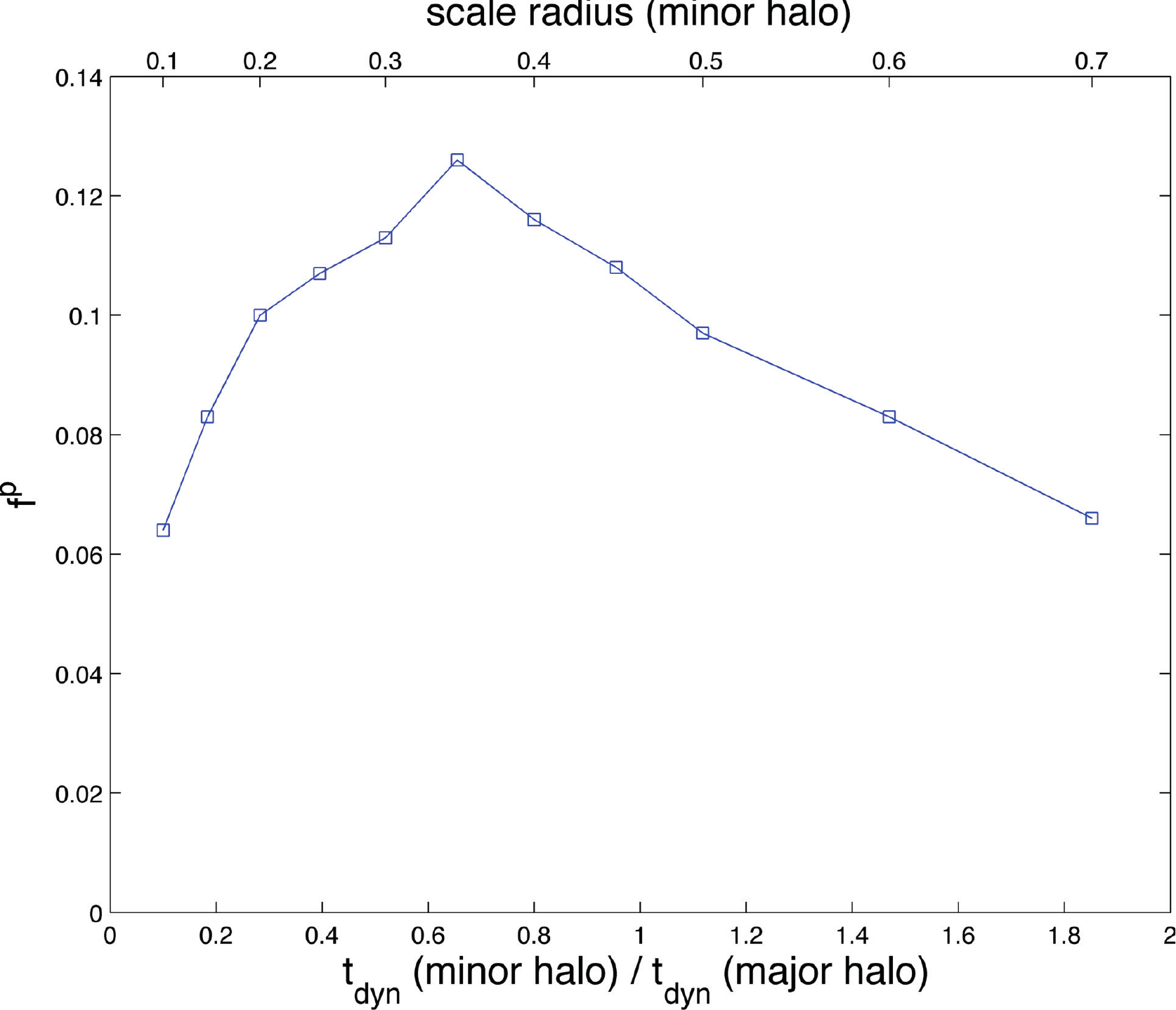}
\caption{The fraction of particles initially in the small structure which
are ejected , $f^p$, as a function of the ratio between the dynamical times
of the small structure and the big structure.}
\label{fig:fpt}
\end{figure}

In figure \ref{fig:fpt} we plot the fraction of the minor halo
particles that get ejected for the different values of the minor halo's
scale radius. We see that the number of ejected particles grows with 
the dynamical time up to a peak that corresponds roughly to a small 
structure with a dynamical time $t_{\text{dyn}}(\text{small}) \simeq 0.7 t_{\text{dyn}}(\text{big}) $,
and then beyond this point the number decreases monotonically again.

This behaviour is expected from the considerations above.  Basically we 
need to compare two timescales, namely the time it takes the small 
structure to cross the big structure, and the time for a
typical orbit in the small structure.  The first timescale corresponds
to the crossing time for the big structure, which is proportional to
our definition of the dynamical time, $\tau_\text{dyn} = 4r_\text{s}
/v_c(4r_\text{s})$.  The second timescale is similar, but defined for the
small structure.

Thus, for a very compact small structure, the particles in the small
structure will make many orbits while crossing the big structure, and
fewer receive sufficient increase in energy to leave the structure. On
the other hand, for a very dilute small structure, the particles in
the small structure will perform much less than one orbit while
crossing the big structure, rendering the motion almost adiabatic.
An extrapolation of this finding is that smooth accretion should not
lead to particle ejection.

\section{Tracing particles in an analytical potential}
\label{sec:MFE}

In order to demonstrate that the ejection of particles during mergers
is a mean field effect, we implement a toy-model in which we consider
one particle moving in an analytical time-dependent potential 
approximating that of the merging structures. In this
way we eliminate dynamical friction entirely. We also eliminate any two-body effects 
which may be a possible spurious source of particle ejection.

\subsection{Simulation Set-Up}
\label{sub:setup2}

We simulate a merger with the same characteristics as the one
described in section \ref{sub:example}, but now 
using a smooth potential instead of the $N$-body approach used earlier.
We then follow the orbits in this potential of a test particle which 
initially is bound to the small structure\footnote{This approach is similar to that of \citep{teyssier}, but differs
in that the latter considered the sum of particles on large orbits
 (wandering stars) and truly ejected particles. Since we here
 consider separately the ejected particles (and don't consider the
 particles merely on large orbits), a direct comparison is not
 possible.}.
For both structures we use 
analytical potentials for the same Hernquist model 
\citep{1990ApJ...356..359H}
\begin{equation}
\Phi(r) = - \frac{GM/r_\text{s}}{1 + r/r_\text{s}} \, ,
\end{equation}
where the large structure has a scale radius $r_{s1} = 1$ and mass
$M_1=1$, and the smaller one has $r_{s2}= 0.3$ and $M_2=0.1$. 
The two potentials are initially at rest and centered at $15 r_{s1}$ 
apart along the $x$-axis, which is the turnaround radius of the big 
structure, just as discussed in section \ref{sub:setup1}. For each time 
step the potential
in which the test particles move is then evolved as follows.
The large structure is kept fixed at all times, while the centre 
of the small structure follows the free fall trajectory induced by 
the large large structure until the time when 
their centres coincide. For the subsequent evolution 
we consider four extreme cases for the interaction between 
the two potentials, representing extremes between which the   
full $N$-body simulation would most likely be.

In the first set of simulations (Case 1) the small structure stops
instantaneously when its centre overlaps the centre of the bigger
structure; in the second set (Case 2) we let the small structure
continue its motion in the $x$ direction, following still its
free fall trajectory due to the gravity of the large structure.

In the third and fourth cases, we do as for the first
and second cases respectively, but now include a small 
impact parameter in order to model non-head-on collisions.  
We achieve this by shifting the velocity direction by hand
when the small halo is at $x = 5$.

In each simulation we follow the motion of one particle
initially bound to the small potential.  We choose its initial
position and velocity similarly to how this was done in setting
up the $N$-body simulations, with its radius sampled from 
the Hernquist mass profile. We assign an initial velocity by determining the
typical speed at the radius just chosen, using an analytical
expression for the radial component of the velocity dispersion.  We
then let the system evolve. For each time step the gravitational
attraction of the big structure on the small one, and of both
structures on the particle, are calculated using a leapfrog
integrator.

\subsection{Particle trajectories}
\label{sub:results}

\begin{figure}[p!] 
\begin{minipage}[b]{0.47\linewidth}
	\centering
	\includegraphics[width=0.95\linewidth]{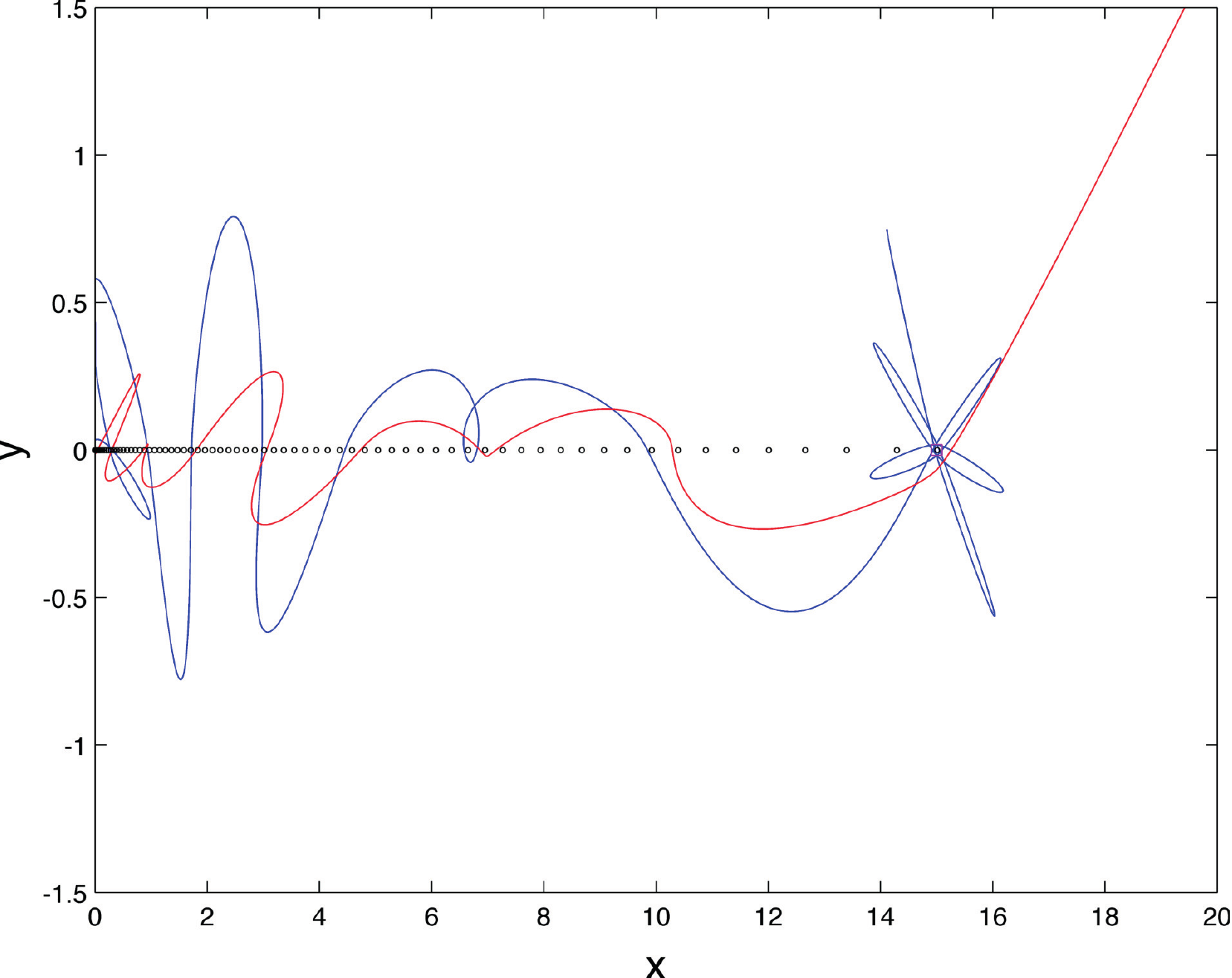}
	\caption{ Positions of the centre of the two potentials, and 
	orbits of two particles projected in the spatial coordinates $(x,y)$.
	Case 1 is a head-on collision where the small potential
	    stops instantaneously as it reaches the big potential, which stays fixed at $x =
	    15$ and $y=0$ (magenta square).  Black dots are equal time
	    intervals of the positions in $x$ and $y$ of the centre of the
	    small potential.  The blue orbit shows a particle that happens to
	    be bound in the final resulting potential, and the red orbit
	    shows a particle that gets ejected from the system.}
	\label{fig:orbits1}
	
  \end{minipage} 
\hfill
  \begin{minipage}[b]{0.47\linewidth}
	
	\centering
	\includegraphics[width=0.95\linewidth]{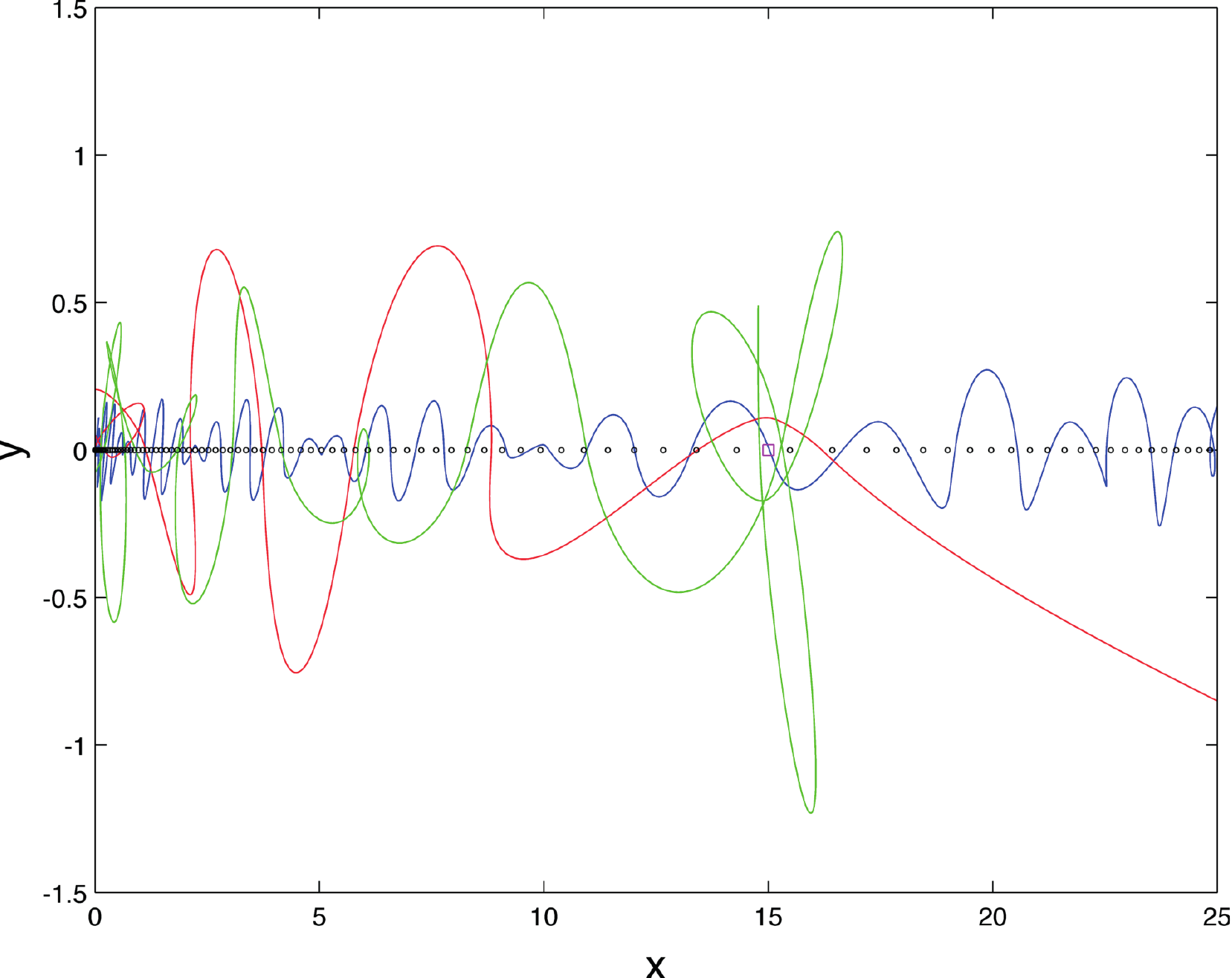}
	\caption{Case 2 is a head-on collision where the small potential
	  continues moving after passing the centre of the big structure.
	  Black dots are equal time intervals of the positions in $x$ and $y$
	  of the centre of the small potential.  The blue orbit shows a
	  particle that happens to be caught in the final resulting potential,
	  and the red orbit shows a particle that gets ejected from the
	  system.  The green orbit shows a particle that gets bounded to the
	  larger potential after the merger.\vspace*{0.4cm}}
	\label{fig:orbits2}
    \end{minipage}

\end{figure}

\begin{figure}[p!] 

  \begin{minipage}[b]{0.47\linewidth}

	\includegraphics[width=0.95\linewidth]{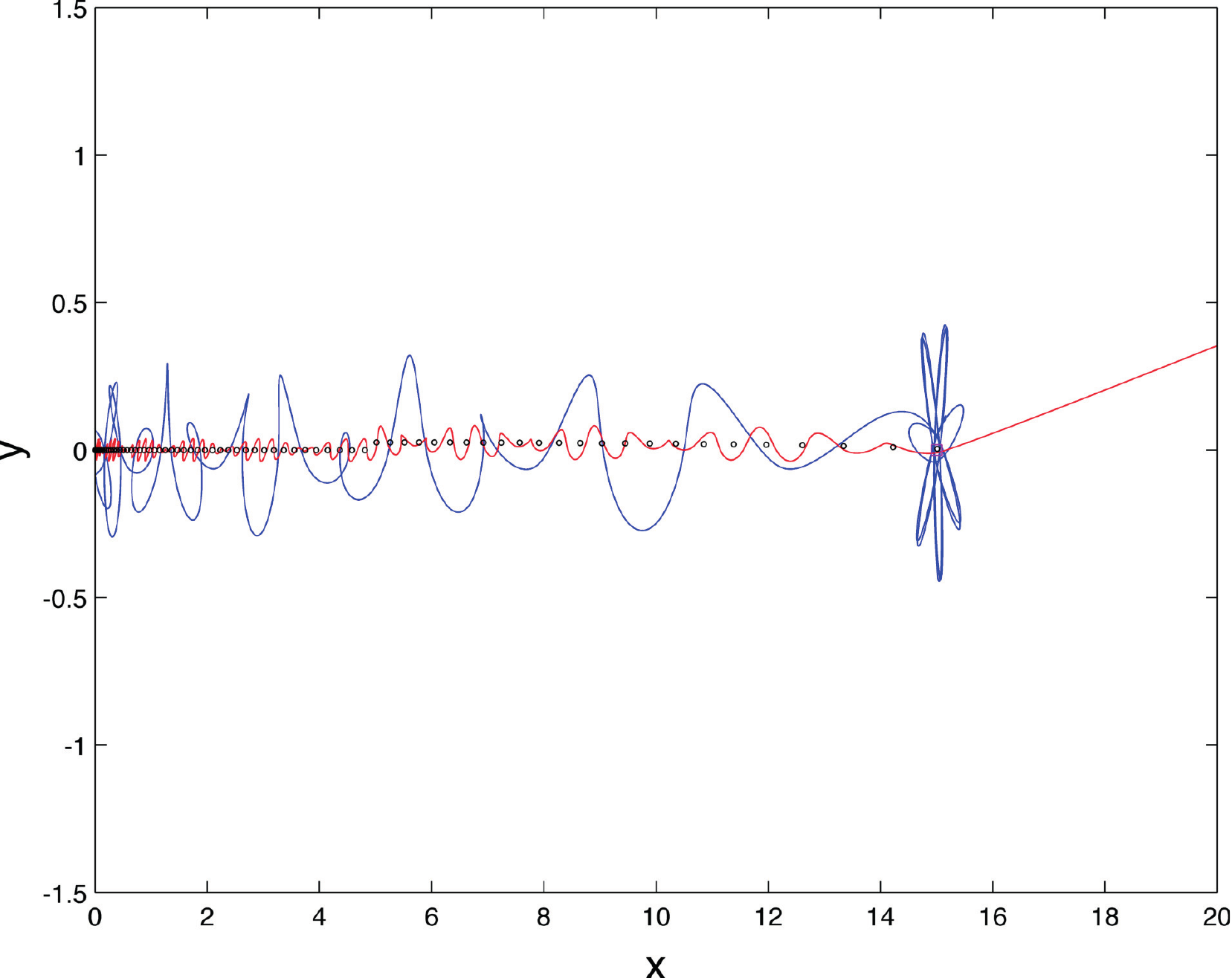}
	\caption{ Case 3 is a collision with non-zero impact parameter, where
	  the small potential stops as it reaches the big potential, which
	  stays fixed at $x = 15$ and $y=0$ (magenta square).  Black dots are
	  equal time intervals of the positions in $x$ and $y$ of the centre
	  of the small potential.  The blue orbit shows a particle that
	  happens to be caught in the final resulting potential, and the red
	  orbit shows a particle that gets ejected from the system.\vspace*{0.5cm}}
	\label{fig:orbits3}

  \end{minipage}
  \hfill
  \begin{minipage}[b]{0.47\linewidth}

	\centering
	\includegraphics[width=0.95\linewidth]{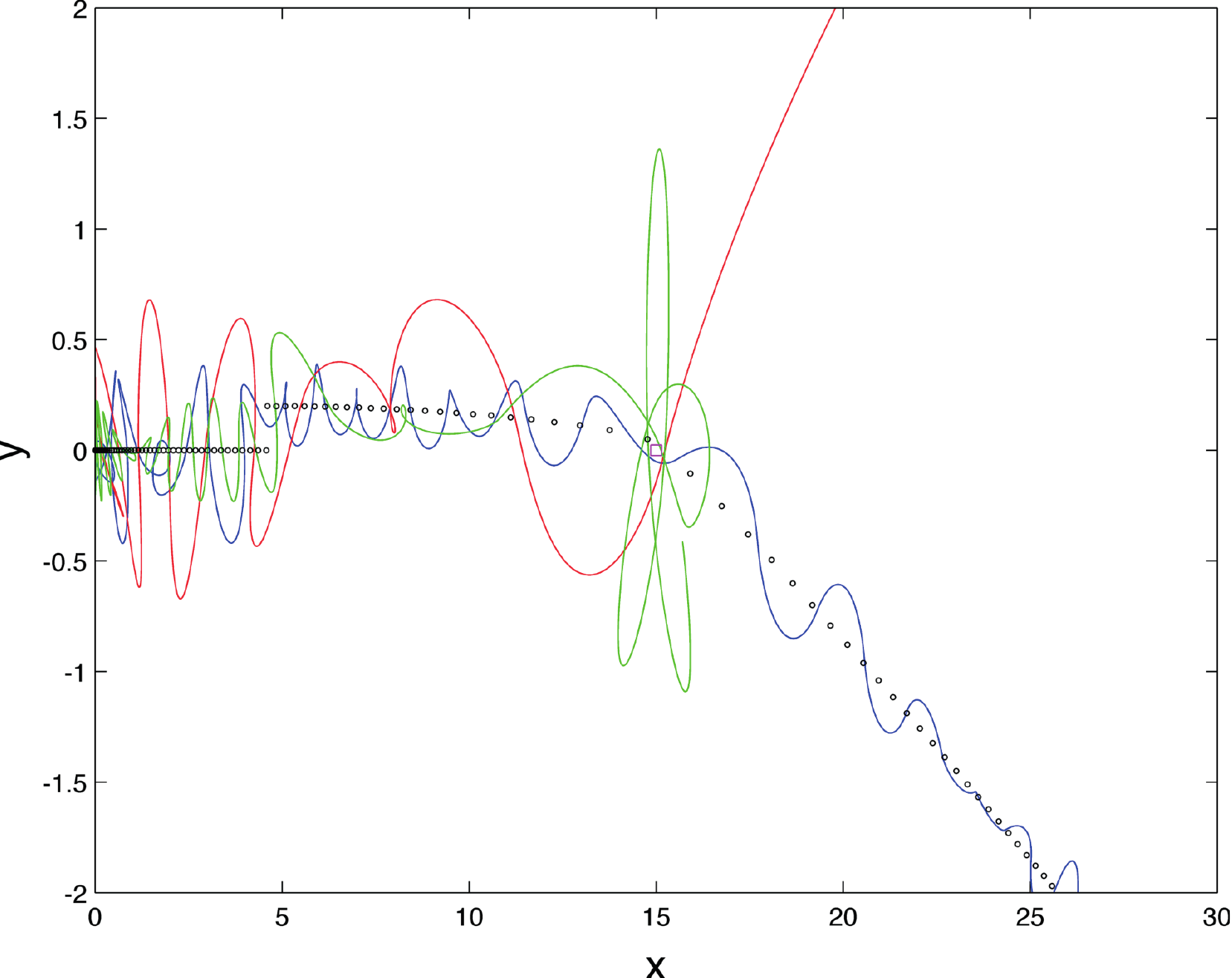}
	\caption{ Case 4 is a collision with non-zero impact parameter, where
	  the small potential continues moving after passing the centre of the
	  big structure.  Black dots are equal time intervals of the positions
	  in $x$ and $y$ of the centre of the small potential.  The blue orbit
	  shows a particle that happens to be caught in the final resulting
	  potential, and the red orbit shows a particle that gets ejected from
	  the system.  The green orbit shows a particle that gets bounded to
	  the major potential after the merger.}
	\label{fig:orbits4}

  \end{minipage} 
\end{figure}

In figure \ref{fig:orbits1} we plot the orbits of two different
particles, hence two different simulations, belonging to the first set
of simulations, 
in which the potential of the small structure stops (forcing its velocity to zero) when its
centre reaches that of the large structure. 
The red particle is ejected from the system, the blue one
remains bounded in the resulting structure and starts orbiting around
it. 

In figure
\ref{fig:orbits2}-\ref{fig:orbits4}
we present similar plots for the other three cases, picking for 
each case a particle that
remains bounded and another that gets ejected.

To get statistics on the particle ejection, we have run $1000$ simulations 
for each case, choosing the test particle randomly as described above.
To check for ejection  we control the
sign and constancy of the total energy of the particle. In the second
and in the fourth cases, when the small potential keeps moving in free fall after passing through the big one, it can
happen that the particle neither follows the small potential nor
gets ejected, but it gets trapped by the bigger potential. If this is
the case, the kinetic energy of the particle has to be calculated with
respect to the potential of the large structure. We show examples in figures
\ref{fig:orbits2} and \ref{fig:orbits4} (in green) of 
orbits of particles that get bound to the big potential after the
collision.
\clearpage
\subsection{Fraction of ejected particles}

We summarise the results of the $4000$ simulations in table
\ref{tab:stat}.  For the simulation with the same
parameters calculated with the $N$-body code and described in section
\ref{sub:example}, the percentage of the ejected particles is $11\%$,
which lies between the two values of cases $1$ and $2$.

We notice that in the case of non-head-on collisions the fraction
of the mass ejected increases slightly.

\begin{table}
  \centering
  \caption{\label{tab:stat} Statistics for particle ejection. $1000$ simulations were run for each case. In cases 1 and 3, after the potentials of the small and large structures overlap, the simulation continues only for the test particle, leaving the potentials fixed. In cases 2 and 4 the small potential continues through the large potential, and hence the particle may end up in either of the two, or be ejected.}

\begin{tabular}{ lccc}
& Ejected &	Small& Big\\
&  &	 halo & halo\\ 
\hline
 {\bf Case 1} &  & & \\	
 Halo stops. & $14 \%$&$86 \%$ &  \\					
Head-on. & & &  \\	
 & & &  \\				
  {\bf Case 2}&  & & \\	
Halo doesn't stop. & $5.3\%$&$82.4 \%$ &$12.3\%$  \\		
     Head-on. & & &  \\		
      & & &  \\				
{\bf Case 3} &  & & \\	
Halo stops. & $17 \%$&$83\%$ &  \\	
 Impact parameter. & & &  \\	
  & & &  \\				
 {\bf Case 4}&  & & \\	
Halo doesn't stop.& $7.2\%$&$76.7\%$ &$16.1\%$  \\	 
  Impact parameter. & & &  \\	
\end{tabular}
\end{table}

\section{The relation between scale radius and fraction of ejected particles}

We will now study how the fraction of ejected particles from the small structure depends on the size of the minor halo in the merger, and compare the dependence in the $N$-body simulations with the dependence in simulations, where individual particles are tracked in analytic potentials.

\begin{figure}
\centering
\includegraphics[width=0.8\textwidth]{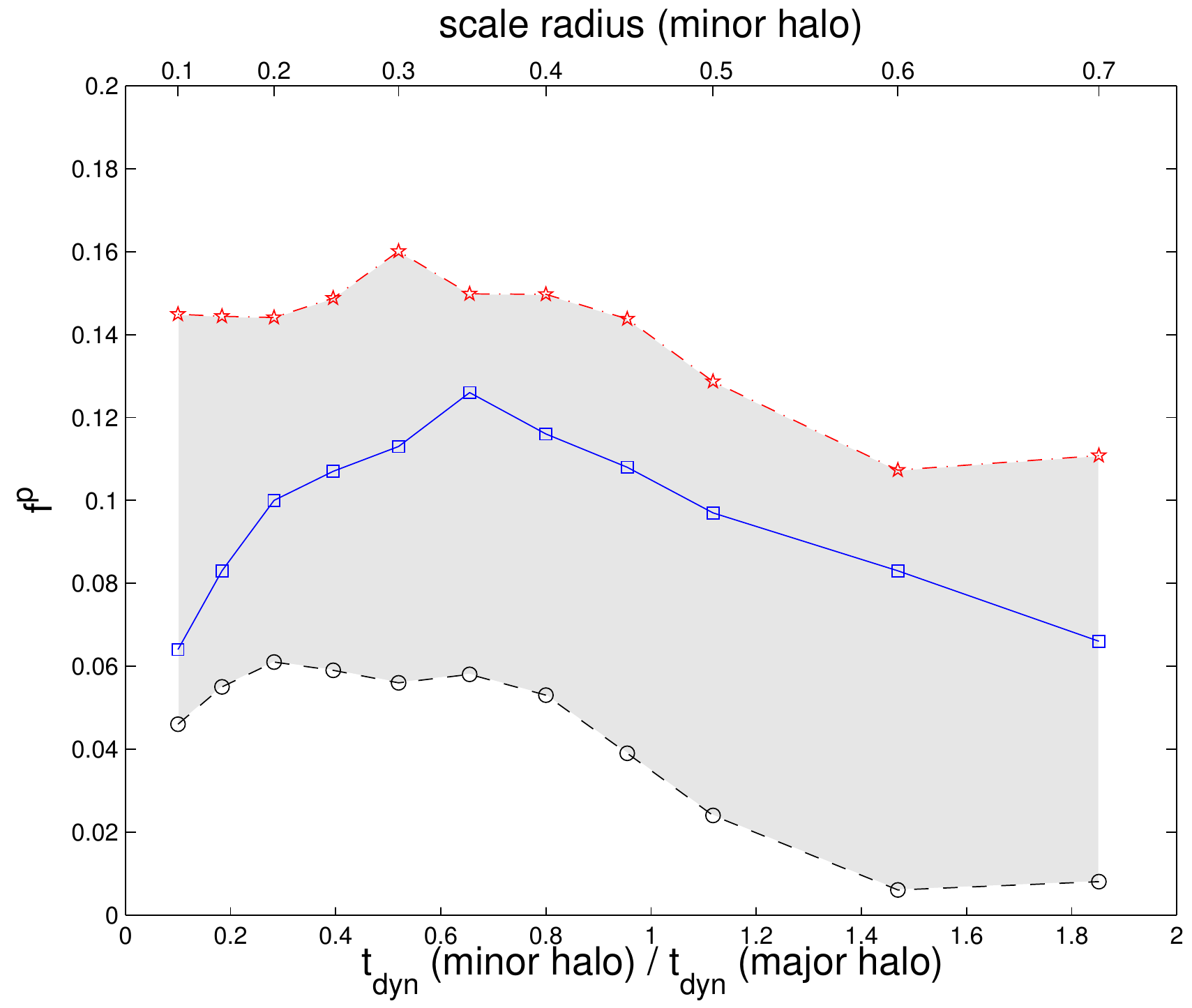}
\caption{The fraction of ejected particles for the head-on mergers for the $N$-body simulations (blue), the analytical potential model where the minor halo is stopped when reaching the big one (red dotted), and analytical model where the minor halo continues with free fall velocity through the major halo (black dashed).}
\label{fig:fp_vs_rs}
\end{figure}

Figure \ref{fig:fp_vs_rs} shows the ejected fraction for the head-on mergers for the $N$-body simulations (from section~\ref{sub:nbody}) and the analytical models, where we track particles in an analytic potential (from section~\ref{sec:MFE}).

The first thing we notice is that the fraction of ejected particles in the $N$-body simulations falls within the range spanned by the two analytical models. This result is independent of the scale radius of the minor halo. It is not a surprising result since the way the minor halos evolves in the $N$-body simulation falls somewhere in between the extreme cases studied in the analytical models, where the minor halo potential is either totally unaffected by the major halo, or it is stopped instantly when it reaches the center of the major halo.

Another result is that the $N$-body simulation and both analytical models all show a decline in the fraction of ejected particles when the minor halos scale radius is larger than $r_\text{s}\simeq 0.35$ (i.e. 35\% of the major halos scale radius). The similarities (the decline in fraction of ejected particles for  $r_\text{s}>0.35$) and differences (i.e. the different overall normalization in the different models) between the analytical models and the $N$-body simulations suggest several conclusions

\begin{itemize}
	\item The detailed destruction history of the minor halo is important for the overall normalisation of the fraction of ejected particles.
	\item In the setups studied in this work, there is a decline in fraction of ejected particles when the minor halos scale radius reaches 35\% of the scale radius of the major halo. This results holds when the mass ratio of the two halos are fixed at 1:10, and will likely change for different mass ratios.
	\item In the models, where particles are tracked in an analytical potential, dynamical friction between the two merging halos is not present. Because of the similarities between these models and the $N$-body simulations, we conclude that it is possible to eject particles without dynamical friction. It is still possible that dynamical friction can affect the exact number of ejected particles, but we have shown that it is not the main driver of ejection in minor mergers.
\end{itemize}

\section{Discussion and Conclusion}

We have shown that during minor mergers approximately $5-15\%$ of the particles from the minor halo are ejected, making the phenomenon quantitatively important in structure formation scenarios. In analytical models (with dynamical friction turned off) and $N$-body simulations, there exists similar relations between the fraction of ejected particles as a function of the ratio between the scale radii of the two halos in the merger. This similarity shows that dynamical friction is not the driving mechanisms for the ejection of particles.

Instead we find that the relevant mechanism is the increase in the total energy of individual particles arising from the time-dependence of the mean field potential during the merger process.  The ejected particles are those that travel in a deep potential as they move in towards the plane of the center of the main halo, and a shallower potential as they move out. In a minor merger these are the particles which originate in the small halo and cross the plane of the center of the large halo slightly later than the center of the halo they originally are bound to does.

In this study we have only considered collisionless
  systems. It is unclear how the presence of baryons will affect the
 number of ejected particles. It has been proposed that the potential
 variations due to supernova feedback can turn galaxy cusps into
 cores \citep{Pontzen2013,Amorisco2013}, and it is 
likely that these potential changes can also affect the number of
particles ejected during a merger. It is expected that the merging
 of two galaxies leads to a significant increase in the star
formation rate \citep{Hayward2013}, so an increase in the potential
variation due to feedback processes is expected during a
merger. The exact role of feedback processes will of course depend
on the actual feedback model used in the simulation (many different
feedback models exist, e.g. \citep{Agertz2013,Vogelsberger2013,Stinson2013,Marinacci2014}).

Our finding provides an explanation for the origin of high-velocity
component of dark matter particles observed in cosmological $N$-body
simulations.  This component of high-velocity particles is important
since it potentially may give a clear signature in underground dark
matter detectors~\citep{2012arXiv1208.0334B}. This is because
particles ejected throughout the merging history of the
Universe should permeate space, and also be present in the
Earth's neighbourhood, at energies significantly higher than the
  equilibrated dark matter component.  Thus, even though the ejected particles
only contribute around $3\%$ of the dark matter near Earth, then they could
still induce a peaked signal on top of the broad bump from 
the thermalized dark matter component.

\acknowledgments 
The Dark Cosmology Centre is funded by the Danish National Research Foundation.
MJ thanks the Dark Cosmology Centre for financial support as a visitor at the
centre.

\def\aj{AJ}
\def\araa{ARA\&A}
\def\apj{ApJ}
\def\apjl{ApJ}
\def\apjs{ApJS}
\def\apss{Ap\&SS}
\def\aap{A\&A}
\def\aapr{A\&A~Rev.}
\def\aaps{A\&AS}
\def\mnras{MNRAS}
\def\na{New Astronomy}
\def\jcap{JCAP}
\def\nat{Nature}
\def\pasp{PASP}
\def\aplett{Astrophys.~Lett.}
\def\physrep{Physical Reviews}
\bibliographystyle{JHEP}
%\bibliography{ref}

\providecommand{\href}[2]{#2}\begingroup\raggedright\endgroup

\end{document}